\documentclass[12pt,sort&compress]{article}

\usepackage[dvips]{color}
\usepackage{amsmath}
\usepackage{graphicx} 
\usepackage{graphicx}
\usepackage{tabularx}
\usepackage{subfig}
\usepackage{cite}
\usepackage{amsfonts}
\usepackage{tabularx,ragged2e}
\usepackage{xcolor}
\newcolumntype{C}{>{\Centering\arraybackslash}X} 
\usepackage[outercaption]{sidecap}    
\usepackage{hyperref}

\newcommand{\erf}{\mathrm{erf}}

\begin{document} 
\title{Entropy barriers and accelerated relaxation under resetting}
\date{}
\author{Pascal Grange\\
 Xi'an Jiaotong-Liverpool University\\
111 Ren'ai Rd, 215123 Suzhou, China\\
\normalsize{{\ttfamily{pascal.grange@xjtlu.edu.cn}}}}
\maketitle

\begin{abstract}

The zero-temperature limit of the backgammon model under resetting is studied.
 The model is a balls-in-boxes model whose relaxation dynamics is governed by the density of boxes containing just one 
 particle. As these boxes become rare at large times, the model presents an entropy barrier. 
  As a preliminary step, a related model with faster relaxation, known to be mapped to a symmetric random walk, 
 is studied by mapping recent results on diffusion with resetting onto the balls-in-boxes problem. 
  Diffusion with an absorbing target at the origin (and  diffusion constant equal to one), 
 stochastically reset to the unit position, 
  is a continuum approximation to the dynamics of the balls-in-boxes model, with resetting to a configuration maximising the 
 number of boxes containing just one ball.
 In the limit of a large system, the relaxation time of the balls-in-boxes model under resetting  is finite. 
  The backgammon model subject to a constant resetting rate is then studied using an adiabatic approximation.

\end{abstract}

\tableofcontents

\section{Introduction}

Out of-equilibrium physics has recently given rise to  remarkable results 
 on the dynamics of systems subject to resetting.
  The corresponding renewal equations  \cite{evans2011diffusion,evans2011optimal,evans2018run} have a wide range of applicability 
 and have yielded predictions in models of active matter\cite{evans2018run,refractory},  predator-prey dynamics \cite{mercado2018lotka,toledo2019predator}, population dynamics \cite{ZRPSS,ZRPResetting},
 as well as stochastic processes \cite{lapeyre2019stochastic,gupta2018stochastic,basu2019long,basu2019symmetric} (see \cite{topical} for a recent review, and references therein for more applications).  In particular, the expected time 
 for a diffusive random walker in one dimension to reach an absorbing target, which is infinite in the ordinary case, 
 is finite if the walker is stochastically reset to its initial position, at exponentially-distributed times \cite{MajumdarDiffusion}.\\

On the other hand, glassy systems have inspired models with very slow relaxation to 
 equilibrium. The first explicit example of these models, termed the backgammon model, involves 
 a fixed number of distinguishable particles distributed amongst a fixed number of boxes \cite{ritort1995glassiness,franz1995dynamical,backgammon}.
  At each time step (in the zero-temperature version of the model), a particle is drawn uniformly and put in another ({\emph{non-empty}}) box. If the energy is defined as 
 the {{negative}} of the number of empty boxes, there is no energy barrier to the relaxation of the model, 
 as the energy can only go down.
 However, there is an entropy barrier: the configurations leading to a net decrease in energy (configurations in which 
 there are boxes containing just one particle) become increasingly rare during relaxation. 
 The occupation-number probability in non-empty boxes of this model,
  as well as that of a closely related model with indistinguishable particles and a faster
 relaxation, have been mapped to one-dimensional random walks \cite{godreche1995entropy}, in which the position of the random walker plays
 the role of the occupation number.
It is therefore natural to ask how results 
 on the dynamics of random walks under resetting \cite{evans2011diffusion} can induce resetting prescriptions for models 
 with entropy barriers (with computable consequences on the acceleration of the dynamics).\\

The paper is organised as follows.  Section 2 is a review of the mapping of  
  balls-in-boxes models without energy barrier onto random walks with an absorbing trap, 
  introduced in \cite{godreche1995entropy}. This mapping applies both to the zero-temperature version 
 of the backgammon model (termed model A), and to a related model  (termed model B), 
 in which particles are indistinguishable and the dynamics is generated by drawing departure boxes uniformly 
  amongst non-empty boxes. Model B enjoys both faster relaxation and easier solution than model A, 
 because it  maps to a symmetric random walk (up to a density-dependent redefinition of time). 
  In Section 3 a resetting prescription is defined, and the results of \cite{evans2011diffusion} on one-dimensional diffusion subject to resetting are mapped back 
 to model B. This allows, in the limit of large systems,
  to calculate the relaxation time of model B under resetting.
 In Section 4 the resetting prescription is adapted to model A. The relaxation of model A under resetting 
 is studied in the adiabatic approximation, using a functional expression of the 
  generating function of the occupation-number probability. In this approximation, 
 the relaxation dynamics of the energy is expressed 
 in closed form in terms of the Laplace transform of the probability of occupation-number one, 
 taken at the resetting rate. 
 This prediction for the time evolution of the energy is compared to direct Monte Carlo simulations. \\

\section{Description of the models}

\subsection{Balls-in-boxes models with entropy barriers }
Consider balls-in-boxes models, in which  the energy of the system
  is defined as the {{negative}}  of the density of empty boxes (this density is the number of empty boxes divided by the total number $N$
 of particles in the system). 
  If the dynamics of a model does not allow moves  that put a particle into an empty box, the energy can only decrease.\\

 The zero-temperature version of the backgammon model \cite{ritort1995glassiness} (called model A in \cite{godreche1995entropy}) is defined as follows. There are $N$  distinguishable particles distributed amongst $M$ boxes.
 At each time step a particle is randomly drawn (uniformly), 
 and put into another box, randomly drawn (uniformly) from the set of the other non-empty boxes. 
    No  change of configuration involves putting a particle into an empty box{{;}}  in this sense the
 system has no energy barrier. However, 
 as time increases, particles tend to be taken from  boxes containing large numbers of particles. Such moves do not lower the energy,
 hence the dynamics becomes very slow. Relaxation has an entropy {{barrier}} because moves lowering the energy (moves in which the drawn 
 particle is alone in its box before moving) become very rare.\\

 Consider the following related model (called model B in \cite{godreche1995entropy}). There are still  $N$ particles distributed amongst $M$ boxes, but the particles are  indistinguishable.
  At each time step, a departure box is  randomly drawn (uniformly) 
 from the set of non-empty boxes. A destination box is drawn uniformly 
 among the other non-empty boxes. A particle is taken from the departure box and put into the destination box. As departure boxes are drawn uniformly, boxes
 containing just one particle are as likely as boxes with more particles to lose one particle. The relaxation in model B is therefore faster than in model A.\\

 For definiteness let us consider in both models the case $N=M$. 
{{The minimum energy of the model in the limit of large systems is therefore $-1$.}} 
 {{Let us}} choose the initial configuration with one 
 particle in each box. In this configuration, every possible move decrases the energy  in both models. 
 To accelerate the dynamics by resetting the configuration of the system, we will have to pick a resetting configuration 
  {{that is connected by the dynamics to the maximum possible number of states compatible with a given level of energy. 
 This will be achieved by putting exactly one particle in each of the non-empty boxes (except one box which receives the rest of the particles).}}
 The mapping of model B 
  to a one-dimensional diffusive  random walk with an absorbing target
  will allow us to map the solution of the one-dimensional diffusion process under resetting
 to the initial position, to a solvable resetting  prescription for the balls-in-boxes model.

\subsection{Mapping balls-in-boxes models to  random walks with an absorbing point at the origin}

 Let us review the derivation of \cite{godreche1995entropy}. An active box is 
defined as a non-empty box. The number of active boxes at time $\tau$ is denoted by $M(\tau)$.
 For any integer $k$, let us denote by $n_k(\tau)$ the number of active boxes at time $\tau$ containing $k$ 
  particles. In these notations we can write:
\begin{equation}\label{notDef}
 \sum_{k\geq 1} n_k(\tau) = M(\tau),\;\;\;\;\;\;\;\sum_{k\geq 1} k n_k(\tau) = N.
\end{equation}
As empty boxes are not active boxes, the notation implies
\begin{equation}
n_0(\tau) = 0.
\end{equation}
 To denote the number of empty boxes, let us introduce the extra notation $n_0^\ast(\tau)$.
 The energy $E(\tau)$ of the system at time $\tau$ is defined as the {{negative}}  of the density of empty boxes:
\begin{equation}\label{defEnergy}
 E(\tau):=-\frac{n_0^\ast(\tau)}{N}.
\end{equation}
At time $\tau$, the $N$ particles are distributed amongst the $M(\tau)$ active boxes.
Consider  the mean density in the active boxes at time $\tau$, denoted by $\lambda(\tau)$:
\begin{equation}
\lambda(\tau) :=  \frac{N}{M(\tau)}\geq 1. 
\end{equation}
 The lower bound is saturated if there is just  one particle in each box.\\

 Consider the  density $f_k$ of active boxes at occupation number $k>0$, and introduce the 
 natural notation $f_0^\ast$ for the density of empty boxes: 
 \begin{equation}
 f_k(\tau) :=  \frac{n_k(\tau)}{N},\;\;\;\;\;\;\;f_0^\ast(\tau):=\frac{n_0^\ast(\tau)}{N},\;\;\;\;\;\;\;\; f_0(\tau) =0. 
\end{equation}
 From now on let us consider the case where there are as many particles as boxes, and the initial configuration has
 maximal energy:
 \begin{equation}\label{fromEq6}
 M=N,\;\;\;\;\;\; n_k(0) = \delta_{k,1} \times N.
\end{equation}
These notations induce the following normalisation condition:
 \begin{equation}\label{normalisations}
 \sum_{k\geq 1} f_k(\tau) =  \frac{M(\tau)}{N} = \frac{1}{\lambda(\tau)} = \frac{M-n_0^\ast(\tau)}{N} = 1 - f_0^\ast(\tau).
\end{equation}
Using the definition of the energy in Eq. \ref{defEnergy}, we relate the density 
  in active boxes $\lambda(\tau)$ to the energy of the system:
\begin{equation}\label{defaultMin}
 \frac{1}{\lambda(\tau)} =  1 + E(\tau).
\end{equation}

\section{Model B under resetting}

\subsection{Mapping model B to a symmetric random walk}
 The following master equation for the density profile of active boxes,
 worked out in  \cite{godreche1995entropy}, describes the dynamics of model B:
\begin{equation}\label{master}
\begin{split}
 \frac{d f_k}{d\tau}(\tau) &= \lambda(\tau)\left( f_{k+1}(\tau) -2 f_k(\tau) + f_{k-1}(\tau) \right),\;\;\; \forall k >0.\\
 f_0(\tau) &= 0.
\end{split}
\end{equation}
{{The first term corresponds to processes that decrease the number of particles in a box with $k+1$ particles (the rate at which  a box with $k+1$  particles is chosen as  departure box
 equals $n_{k+1}(\tau)/M(\tau) = \lambda(\tau) f_{k+1}(\tau)$). The second term corresponds to processes that decrease the number of particles in a box with $k$ particles, and to processes that increase the number of particles in a box with $k$ particles. The third term  corresponds to processes that increase the number of particles in a box with $k-1$ particles (the rate at which a box with $k-1$  particles is chosen as  destination box
 equals $n_{k-1}(\tau)/M(\tau) = \lambda(\tau) f_{k-1}(\tau)$, even in the case $k=1$, because $f_0=0$ and empty boxes are not departure boxes).}\\

As the  mean density  in the active boxes is positive, we can define a new time variable $t$ by the differential 
 relation:
\begin{equation}\label{reparamTime}
 dt := \lambda(\tau) d\tau,
\end{equation}
 {{which induces an expression of the original time as a function of the new time variable
 (see Eq. \ref{integrating} for the explicit calculation of $\tau(t)$ in the system subjected to resetting).
 Let us  reparametrise the occupation-number probability by:
 \begin{equation}
  g_k(t) := f_k\left[\tau(t)\right],\;\;\;\forall k \geq 0.
\end{equation}}}
  The time evolution described by Eq. \ref{master}  becomes a symmetric random walk with an absorbing trap
 at the origin:
 \begin{equation}\label{randomWalk}
\begin{split}
   \frac{d g_k}{d t}(t) &= g_{k+1}(t) -2 g_k(t) + g_{k-1}(t),\;\;\;\;\;\forall k >0,\\
 {{ g_0(t)}} &= 0.
 \end{split}
\end{equation}
 Moreover, the index $k$ (which denotes the number of particles in the active boxes) 
  plays the role of the position of the random walker.\\

 {{In the limit of an infinite number of particles, the state of the system is described by an infinite family of functions $(g_k(t))_{k\in \mathbf{N}}$.}}
 Let us approximate the model by a continuum, as proposed in \cite{godreche1995entropy}, where the discrete variable $k$ become a positive variable $h$.
  The family of functions $(g_k(t))_{k\in \mathbf{N}}$ becomes a function of two variables $g(h,t)$, which satisfies 
 the heat equation with an absorbing boundary condition {{at the origin}}:
 \begin{equation}\label{continuumHeat}
\begin{split}
 \frac{\partial g}{\partial t} &= \frac{\partial^2 g}{\partial h^2},\\
g(0,t) = 0 \;\;\;\;\;\forall t,&\;\;\;\;\;\;\;\; g(h',0)=\delta(h' -1)\;\;\;\;\;\forall h'.
\end{split}
\end{equation}
 {{The densities of boxes at integer positions are local  averages of a fluctuating quantity supported on a
 continuum of positions, rather than on integer occupation numbers.}}
 Moreover, integrating over the position variable $h$ yields the continuous analogue of Eq. \ref{normalisations}.
  The  inverse of the density in active boxes (parametrised by the 
 new time variable $t$)  is therefore mapped to the survival {{probability}} of the random walker 
 until time $t$. 
 The propagator for the continuum random walk with an absorbing trap at the origin (conditional on the position $h'$ at time $t=0$, {{and with  diffusion constant equal to $1$}}),
 denoted by $p(h,t|h',0)$, is known to be expressed as
\begin{equation}\label{propEx}
 p (h,t| h',0) = \frac{1}{\sqrt{4\pi t}} \left[    \exp\left(   -\frac{(h-h')^2}{4t}\right) -   \exp\left(   -\frac{(h+h')^2}{4t}  \right) \right].
\end{equation}
  The continuum approximation to the density of active boxes (as expressed in Eq. \ref{normalisations}) 
 is therefore obtained in integral form:
 \begin{equation}
  \frac{1}{\lambda(t)} = \int_0^\infty dh g(h,t) =\int_0^\infty dh \int_0^\infty dh' p(h,t|h',0) g(h',0) =\int_0^\infty dh  p(h,t|1,0).
\end{equation}

\subsection{Resetting prescription}

 Consider a resetting process of the continuous random walk, which brings the system back to the 
 initial configuration, at random times (with a constant rate in the time variable $t$, defined by density-dependent rescaling  in Eq. \ref{reparamTime}). 
 The position of the corresponding random walker is reset to $h=1$ at each resetting event.
 {{The analogue of the position of the random walker with an absorbing target at the origin is the occupation number of  boxes,
 and the analogue of the survival probability is the quantity $\lambda^{-1}$ (the default to the minimum energy, Eq. \ref{defaultMin}). We would like to 
  achieve faster relaxation towards configurations of lower energy. The energy can only decrease if there are boxes containing just one particle. Moreover, the dynamics of the model implies that the energy can only decrease over time.
 We would therefore like
 to define a resetting process in the balls-and-boxes model that does not modify the energy of the system (such a modification would directly advance the  
  relaxation process at each resetting event).   
  From a phase-space perspective, the system is in a certain level set of given energy.
The resetting  
 prescription should  put the system at the boundary of this level set. 
 Such a prescription would lift the 
 entropy barrier by putting the system in a state that is close to as many states of lower energy as possible.}}\\

  {{ The resetting configuration should therefore have as much of the residual probability as possible in the lowest  non-zero occupation number.}} This is achieved by resetting the occupation 
 numbers of all the active boxes (except one) to one. The last active box receives the rest of the particles.
 If the resetting occurs when the system is in a configuration with occupation numbers $(n_0^\ast, n_1,\dots, n_N)$, with $n_0^\ast>0$,
 the configuration  $\overrightarrow{n}^{(R)}$ of the system after resetting reads
\begin{equation}
 \overrightarrow{n}^{(R)} = (  n_0^\ast, n_1^{(R)}, 0,\dots,0, n_{N-n_1^{(R)}} = 1 ),\;\;\;\;\;{\mathrm{with}} \;\;\;\;\; N = n_0^\ast +  n_1^{(R)} + 1.
 \end{equation}

  In this resetting configuration, there are therefore $ n_1^{(R)} = N -  n_0^\ast - 1$ boxes  
 with one particle, and one box with $n_0^\ast + 1$ particles. The number of empty boxes in conserved by the  resetting event,
 as required. The resetting configuration
 therefore depends only on the energy of the system at resetting time (up to a permutation of the boxes, but the model has no spatial structure, so for definiteness we chose the box with more than one particle to be at the end of the array of active boxes). In terms of the occupation-number probability, the resetting configuration corresponds to 
 \begin{equation}\label{resetConfig}
 f_1= 1-\frac{1}{N} -  \frac{n_0^\ast}{N}  = \frac{1}{\lambda} - \frac{1}{N} = \frac{1}{\lambda}\left(  1 - \frac{\lambda}{N}  \right),\;\;\;\;\;\;f_{N- n_0^\ast- 1} = \frac{1}{N},
 \end{equation}
 which sum to the residual probability of survival $1/\lambda$ at resetting time.\\

\subsection{Renewal equation} 

 Let us denote by $1/\lambda^{(r)}(t)$ the survival  probability at time $t$ in the process with resetting (occurring at constant rate $r$ in the time variable $t$),
  and by $1/\lambda^{(0)}(t)$ the (known) survival  probability at time $t$ in the process without resetting. In the time interval $[0,t]$, there has been either no resetting, or at least one resetting. The last of these resetting events happened at time $t-\chi$, for some time $\chi$ in $[0,t]$. 
 These two {{mutually-exclusive}} cases give rise to the two terms on the r.h.s of the following renewal equation \cite{evans2011diffusion,evans2018run}:\\
  \begin{equation}
 \frac{1}{\lambda^{(r)}(t)} = e^{-rt} \frac{1}{\lambda^{(0)}(t)} + r \int_0^t d\chi e^{-r\chi}  Q_0(\chi|R) \frac{1}{\lambda^{(r)}(t-\chi)},
 \end{equation}
 where the symbol $Q_0(\tau|R)$ denotes the probability of survival at time $\tau$  in the symmetric random walk {{(with absorbing target at the origin)}} without resetting,
 conditional on starting at the 
  resetting configuration described by Eq. \ref{resetConfig}:
 \begin{equation}
{{
 Q_0(\chi|R)  = \int_0^\infty dh  \int_0^\infty dk\, p (h,\chi| g,0) \pi_R(k),}}
 \end{equation}
with 
 \begin{equation} 
 {{
  \pi_R(k) =  \left(   1- \frac{\lambda^{(r)}(t-\chi)}{N}     \right)   \delta_1(k)    +  \frac{\lambda^{(r)}(t-\chi)}{N} \delta_{N - n_0^\ast- 1}(k).}}
 \end{equation}
 The propagator $p$ of the process without resetting, expressed in  Eq. \ref{propEx}, gives rise to the relevant survival probabilities 
 between {{resetting}} times: 
\begin{equation}
Q_0(\chi|R) =   \left(   1- \frac{\lambda^{(r)}(t-\chi)}{N}     \right)  \int_0^\infty dh p (h,\chi| 1,0)  
  + \frac{\lambda^{(r)}(t-\chi)}{N}  \int_0^\infty dh p (h,\chi| N - n_0^\ast- 1,0),
 \end{equation}
 \begin{equation}
{{
 \frac{1}{\lambda^{(0)}(\chi)} = \int_0^\infty dh p (h,\chi| 1,0) = \erf\left(\frac{1}{2\sqrt{\chi}} \right).}}
 \end{equation}
The renewal equation for the inverse density in active boxes therefore reads
 \begin{equation}
\begin{split}
 \frac{1}{\lambda^{(r)}(t)}=& e^{-rt} \erf\left(\frac{1}{2\sqrt{t}} \right) \\
  &\;\;+ r \int_0^t d\chi e^{-r\chi} \left[   \left(   1- \frac{\lambda^{(r)}(t-\chi)}{N}     \right)  \erf\left(\frac{1}{2\sqrt{\chi}} \right)   + 
 \frac{\lambda^{(r)}(t-\chi)}{N} \erf\left(\frac{    \frac{N}{\lambda^{(r)}(t-\chi)}- 1   }{2\sqrt{\chi}} \right)    \right] \frac{1}{\lambda^{(r)}(t-\chi)}\\
=&  e^{-rt} \erf\left(\frac{1}{2\sqrt{t}}\right)+ r \int_0^t d\chi e^{-r\chi}  \erf\left(\frac{1}{2\sqrt{\chi}} \right) \frac{1}{\lambda^{(r)}(t-\chi)} \\
  &\;\;+ \frac{r}{N} \int_0^t d\chi e^{-r\chi} \left[ -\erf\left(\frac{1}{2\sqrt{\chi}} \right) +    \erf\left(\frac{    \frac{N}{\lambda^{(r)}(t-\chi)}- 1   }{2\sqrt{\chi}} \right)\right] d\chi.
\end{split}
 \end{equation}

 In the large-$N$ limit the  last term  is negligible and  {{the following renewal equation is therefore satisfied by the inverse of the mean density in the active boxes:}}
 \begin{equation}\label{sameQr}
\frac{1}{\lambda^{(r)}(t)} \simeq_{N\to\infty} e^{-rt} \erf\left(\frac{1}{2\sqrt{t}}\right)+ r \int_0^t d\chi e^{-r\chi}  \erf\left(\frac{1}{2\sqrt{\chi}} \right) \frac{1}{\lambda^{(r)}(t-\chi)}.
 \end{equation}
The above approximation is equivalent to neglecting the finite-size effects due to the presence of one box with more than one particle in the resetting configuration. The large-$N$ approximation was also made when the balls-in-boxes model was mapped to a random walk: in Eq. \ref{randomWalk}, the position  index $k$ was allowed to take any positive value, even though by construction of the model its maximum value is $N$. {{When there is one box with $N$ particles the process ends because there are no other active boxes to move a particle to}.}\\


{{Let us denote by $Q_r(x_0,t)$ the survival probability at time $t$ of a diffusive random walker (in dimension one), 
whose position at time $0$ is $x_0$, in the presence of an absorbing point at position $0$. The diffusion constant is equal to $1$,
 and the random walker undergoes resetting to its initial position $x_0$ at a rate $r$ (this is the notation used in Section 3.1 of the review \cite{topical}). Because of the initial condition and resetting configuration we have chosen in Eq. \ref{continuumHeat}, we will be interested in $Q_r(1,t)$. This survival probability satisfies the following  renewal equation \cite{evans2011diffusion,evans2011optimal}:}}
 \begin{equation}\label{Qr}
Q_r(1,t)  = e^{-rt} \erf\left(\frac{1}{2\sqrt{t}}\right)+ r \int_0^t d\chi e^{-r\chi}  \erf\left(\frac{1}{2\sqrt{\chi}} \right) Q_r(1, t-\chi).
 \end{equation}
{{Moreover, $Q_r(1,0)= 1$ because the initial position of the random walk is fixed and distinct from the origin, and $(\lambda^{(r)}(0))^{-1}=1$
 because the initial density in boxes is $1$, from Eq. \ref{fromEq6}. The quantities $(\lambda^{(r)}(t))^{-1}$ and  $Q_r(1,t)$ are therefore identical in the large-$N$ limit.}} Let us use the results and notations of Section 3 of \cite{topical}. Taking the Laplace transform of both sides of Eq. \ref{Qr} yields
 \begin{equation}
\tilde{Q_r}(1,s)  = \frac{1- \exp\left( -\sqrt{ r+s} \right)}{ s + r \exp\left( -\sqrt{ r+s} \right)}.
 \end{equation}
 Inverting the Laplace transform yields  a parametric representation of the energy of the system in the new time $t$ 
 which for large $t$ goes exponentially 
 fast to the minimum. Let us follow again the notations of \cite{topical}: there exist two quantities $A>0$ and $s_0<0$, that depend on $r$ only, such that 
 {{at large $t$}}
 \begin{equation}\label{paramRep}
 E( t ) = -1 + Q_r( 1, t) = -1 +  A e^{s_0 t}.
 \end{equation}
Moreover, $s_0$ is the solution of 
\begin{equation} 
 s_0 + r \exp\left( -\sqrt{ r + s_0 } \right) = 0,
\end{equation}
 which describes the pole  structure in the Laplace transform.\\

 On the other hand, the original time $\tau$ in the balls-in-boxes model can also be represented in terms of the 
 parameter $t$,  by integrating Eq. \ref{reparamTime} w.r.t. the parameter $t$ {{(and using the fact that $(\lambda^{(r)}(t))^{-1}$ 
  is the survival probability $Q_r( 1,t)$, as we saw from Eqs \ref{sameQr} and \ref{Qr})}}:
\begin{equation}\label{integrating}
\begin{split}
\tau( t ) &= \int_0^t \frac{1}{\lambda^{(r)}(u)}du  \\
 &= \int_0^\infty  Q_r(1, u) du  - \int_t^\infty Q_r(1, u) du \\
 & \underset{ t \to\infty}{\simeq}  \tilde{Q_r}(1,0)+ \frac{A}{s_0} e^{s_0 t}\\
 &= \frac{1}{r}( e^{\sqrt{r}} - 1)+ \frac{A}{s_0} e^{s_0 t}.
\end{split}
\end{equation}
{{Moreover,}} the first term {{$\tilde{Q_r}(1,0)$  equals the mean time to absorption  $\langle T \rangle$ (indeed this mean time is the integral 
 of time $t$ against the negative of the time derivative of the survival probability $Q_r(1,t)$, which is expressed as a Laplace transform upon integration by parts)}}.
 We have to eliminate the parameter $t$ from the two parametric representations {{found in Eqs \ref{paramRep} and \ref{integrating}:}}
 \begin{equation}
1 + E( t )  \underset{ t \to\infty}{\simeq} A e^{s_0 t},\;\;\;\;\;\;\;\;\tau( t )  \underset{ t \to\infty}{\simeq}  \langle T\rangle + A \frac{e^{s_0 t}}{s_0}.
\end{equation}
  Hence the minimum energy  is approached when the  real time gets close to $\tilde{Q_r}(1,0)$:
\begin{equation}\label{affine}
E(\tau ) \simeq  -  1 + s_0( \tau -  \langle T\rangle ),\;\;\;\;\;\;\;\;\tau\to \langle T\rangle^-.
\end{equation}
 The process relaxes therefore  in finite real time if it is reset at a rate proportional to the 
  mean density in the non-empty boxes. {{The constant rate of resetting in the variable $t$ implies that the typical interval of  real time between resets (corresponding to 
 an interval of $r^{-1}$ in the variable $t$) goes to zero towards the end of the relaxation (proportionally to the default to the minimum value):
\begin{equation}
\tau( t + r^{-1}) - \tau(t) \underset{ t \to\infty}{\simeq} -\frac{A}{r} e^{s_0t} \simeq \frac{1}{r}\left( 1 + E[\tau(t)]\right).
\end{equation}
}}

\section{Model A under resetting}

\subsection{The generating function as a functional of the density in active boxes}
Consider the asymmetric random walk corresponding to the 
 zero-temperature version of the backgammon model.  We know  from \cite{godreche1995entropy} that it satisfies the following master equation:
\begin{equation}\label{masterAsym}
\frac{df_k}{d\tau}(\tau) = (k+1) f_{k+1} -(\lambda(\tau) + k )f_k(\tau)  + \lambda( \tau) f_{k-1}(\tau),\;\;\;\;\;\forall k \geq 1.
\end{equation}
{{The first term corresponds to processes that decrease the occupation number in a box with $k+1$ particles. The total rate of these processes is the product of the occupation number in such  boxes (because a particle is picked at every step), by the density of these  boxes. The product $k f_k(\tau)$ is the contribution of processes that decrease the occupation number in a box with $k$ particles, and it comes with a minus sign because 
 these processes decrease the density $f_k$. The term $\lambda( \tau) f_{k-1}(\tau)= n_{k-1}(\tau)/M(\tau)$ corresponds to processes that increase the occupation number in a box with $k-1$ particles: it is the probability that the  destination box of  contains $k-1$ particles. The product $\lambda(\tau) f_k(\tau)$ 
 is the probability that the  destination box  contains $k$ particles. It corresponds to processes that increase the occupation number of a box with $k$ particles, and therefore comes with a minus sign.}}\\

 Again $f_0(\tau)$=0, hence the model is mapped to a one-dimensional random walk with an absorbing target at the origin,  but there is a  density-dependent bias.
 The sum of the above equations over $k$ yields, using the expression  given in Eq. \ref{normalisations} for the mean density $\lambda(\tau)$ in active boxes:
 \begin{equation}\label{densityDynA}
\frac{d}{d\tau}\left( \frac{1}{\lambda(\tau) }\right) = -f_1(\tau).
\end{equation}
The dynamics of the model therefore slows down when the number of boxes with just one particle in them decreases.\\

Consider the following function, which is the sum of the generating function of the process with absorbing 
 condition at $k=0$, and the   density of inactive boxes:
\begin{equation}\label{genAsym}
G(y,\tau) = f_0^\ast(\tau ) + \sum_{k\geq 1 } f_k(\tau) y^k = 1 - \frac{1}{\lambda(\tau)} + \sum_{k\geq 1 } f_k(\tau) y^k.
\end{equation}

 Summing the master equation (Eq. \ref{masterAsym}) over all possible values of $k$, and treating the order-zero term in $y$ 
 separately, we obtain:
\begin{equation}\label{PDEAsym}
\frac{\partial G}{\partial \tau}(y,\tau) = (1-y)\left[ \frac{\partial G}{\partial y}(y,\tau)  -\lambda(\tau) G (y,\tau) + \lambda(\tau) - 1 \right].
\end{equation}
 Let us consider $G$ to be a functional of $\lambda$, and solve the above PDE by the method of characteristics (see \cite{backgammon,ritort1995glassiness,franz1995dynamical,godreche1996long,bialas1997}  
 for an analogous approach at finite temperature).
  The function $\lambda$ can be adjusted by imposing the consistency condition:
 \begin{equation}\label{closure}
 G(0,\tau) = 1 - \frac{1}{\lambda(\tau)}.
\end{equation}
%

 Let us look for a change of variables from $(y,\tau)$ to $(v, \tau)$ that maps the PDE (Eq. \ref{PDEAsym}) to an ODE in the time variable:
\begin{equation}
H( v,\tau) = G( y(v,\tau),\tau), \;\;\;\;\;\;\;\;\;\frac{\partial H}{\partial \tau} = \frac{\partial G}{\partial y}  \frac{\partial y}{\partial \tau} + \frac{\partial G}{\partial \tau}.
\end{equation}
 Matching the combination of  first  derivatives in Eq. \ref{PDEAsym} yields the condition 
\begin{equation}
  \frac{\partial y}{\partial \tau} = y - 1.
\end{equation}
 Integrating this equation, we introduce the variable $v$ as an integration constant and  define the
 change of variables by
\begin{equation}
  y(v,\tau) = 1 + v e^\tau.
\end{equation}
 The resulting ODE reads, with the function $\lambda$ treated as a parameter:
\begin{equation}\label{ODE}
   \frac{\partial H}{\partial \tau} =  ve^\tau \lambda(\tau) H + ve^\tau( 1 - \lambda(\tau)).
\end{equation}
 {{Let us introduce the notation
\begin{equation}
 \rho(\tau) := \int_0^\tau ds e^s \lambda(s).
\end{equation}
The solution of Eq. \ref{ODE} reads}}
 \begin{equation}
\begin{split}
   H(v,\tau ) =& H(v,0) \exp\left(  v \rho(\tau) \right) \\
 &+ \exp\left(  v \rho(\tau) \right)
 \int_0^\tau dq \left[ v( 1-\lambda(q)) e^q \exp\left( -v\rho(q) \right) \right].
\end{split}
\end{equation}
 Transforming back to the variables $(y,\tau)$ yields an expression in which the first term corresponds 
  to the initial condition  (or to the last resetting configuration, as we are going to define it):
\begin{equation}\label{Gfunctional}
\begin{split}
  G(y,\tau) =&\; G( 1 + (y-1) e^{-\tau},0) \exp\left(  (y-1)e^{-\tau} \rho(\tau) \right) +\\
& \exp\left(  (y-1)  e^{-\tau} \rho(\tau)  \right)
 \int_0^\tau dq \left[ (y-1)( 1-\lambda(q)) \exp\left( q-\tau -(y-1) e^{-\tau}\rho( q)  \right) \right].
\end{split}
\end{equation}


\subsection{Adiabatic approximation and resetting prescription}
 At time $\tau = 0$, the boundary condition with just one particle per box reads $G( y, 0 )=y$, and at a resetting time $T$, 
   we would like to accelerate the relaxation of the system to its state of minimum energy. According to Eq. \ref{densityDynA}, 
 we can do so by
 maximising the number of boxes with just one particle in them (without modifying the energy of the system), just as in Eq. \ref{resetConfig}. We therefore 
 impose, at resetting time $T$:
\begin{equation}\label{resetBoundary}
G( y, T) =  1 - \frac{1}{\lambda(T)} + \left( \frac{1}{\lambda(T)} {{-\frac{1}{N}}}\right)y+ \frac{1}{N} y^{N( 1- f_0^\ast) -1} \underset{N>>1}{\simeq} 1 - \frac{1}{\lambda(T)} + \frac{1}{\lambda(T)}y.
\end{equation}
 Again, taking the large-$N$ limit amounts to neglecting the presence of just one box with more than one particle at resetting times. 
  The density $\lambda$ is continuous at  resetting times, because the resetting prescription does not change the 
 number of empty boxes.\\

 {{There are two dynamical time scales 
 in the problem: one is the relaxation of the occupation-number probability to a quasi-stationary (density-dependent) distribution, 
        the other one  is the characteristic time of variation of the density.
 Let us assume that these two time scales decouple and}} that the time $\tau$ in Eq. \ref{Gfunctional} is small compared to the characteristic time of variation of 
 the density $\lambda$. This is 
 the adiabatic approximation that was proposed in the ordinary case \cite{franz1995dynamical}. 
 This approximation 
  should be even more valid 
  in a system with resetting, because each resetting event probes short lengths of time.
  However, we need to take into account the short relaxation time scale towards
  the stationary  distribution of the occupation number $k$ (which can be safely taken as Poissonian with parameter $\lambda(\tau)$ in the ordinary case \cite{franz1995dynamical,godreche1995entropy}). 
   The adiabatic approximation on short time scales {{amounts}} to  substituting the constant density
\begin{equation}
 \Lambda := \lambda(0) \simeq \lambda(\tau)
\end{equation}
to all {{occurrences}} of the density $\lambda$ in Eq. \ref{Gfunctional}. {{We are going to make this approximation 
  (instead of imposing the closure condition of Eq. \ref{closure}).}}
\begin{equation}\label{GLambda}
\begin{split}
  G(y,\tau) =  G( 1 &+ (y-1) e^{-\tau},0) \exp\left(  (y-1)\Lambda \int_0^\tau ds e^{s-\tau} \right) +\\
&  (y-1)( 1-\Lambda ) \int_0^\tau dq  \exp\left( q-\tau + (y-1) \Lambda \int_q^\tau ds e^{s-\tau} \right) \\
=  G( 1 &+ (y-1) e^{-\tau},0) \exp\left(  (y-1)\Lambda (1 - e^{-\tau}) \right) +\\
&   (y-1)( 1-\Lambda ) \int_0^\tau dq   \exp\left( q-\tau + (y-1) \Lambda ( 1-  e^{q-\tau}) \right) .
\end{split}
\end{equation}
 The last  integral can be worked out explicitly:
\begin{equation}
\int_0^\tau dq \exp\left( q-\tau + (y-1) \Lambda ( 1- e^{q-\tau}) \right)  = \exp\left(-\tau +(y-1) \Lambda \right)\times
 \int_0^\tau dq \exp\left( q - (y-1) \Lambda e^{-\tau} e^q \right),
\end{equation}
\begin{equation}
 \int_0^\tau dq  e^q  \exp(- (y-1) \Lambda e^{-\tau} e^q ) = -\frac{e^\tau}{(y-1)\Lambda} \left[ \exp( -(y-1) \Lambda )  - \exp( -(y-1) \Lambda e^{-\tau} ) \right].
\end{equation}
Substituting into the expression of the generating function yields:
\begin{equation}\label{GLambdaEff}
\begin{split}
  G(y,\tau) = & \; G( 1 + (y-1) e^{-\tau},0) \exp\left(  (y-1)\Lambda (1 - e^{-\tau}) \right) \\
& +\left(  1 - \frac{1}{\Lambda}  \right) \left[1 -\exp( (y-1) \Lambda( 1 - e^{-\tau} ) )\right].
\end{split}
\end{equation} 
 Let us assume that the last resetting event happened at time $0$, when the density in active 
  boxes was $\lambda( 0 ) = \Lambda$. The  boundary condition is therefore  given by Eq. \ref{resetBoundary}, which yields: 
\begin{equation}\label{GLambdaEff}
\begin{split}
 G(y,\tau) = &\left( 1 - \frac{e^{-\tau}}{\Lambda} + \frac{e^{-\tau}}{\Lambda} y\right)\exp\left(  (y-1)\Lambda (1 - e^{-\tau}) \right) \\
& + \left(  1 - \frac{1}{\Lambda}  \right) \left[ 1 -\exp\left( (y-1) \Lambda( 1 - e^{-\tau} ) \right) \right]\\
\end{split}
\end{equation}
 Let us introduce the shorthand notation:
\begin{equation}
 L := \Lambda( 1 - e^{-\tau} ).
\end{equation}

 The quantity $L$ depends on $\tau$, which is not reflected in the notation, but it  does not depend on $y$
 and it equals zero at time $\tau = 0$.  
\begin{equation}\label{GLambdaEff}
  G(y,\tau) = \left( 1 - \frac{e^{-\tau}}{\Lambda} + \frac{e^{-\tau}}{\Lambda} y\right) e^{-L} \exp\left( yL \right)+\left(  1 - \frac{1}{\Lambda}  \right) \left(1 - e^{-L}\exp( y L )\right).
\end{equation}
We can extract the term of order $1$ in $y$ and read off $f_1(\tau)$ in the adiabatic approximation:
\begin{equation}\label{f1adiab}
\begin{split}
f_1(\tau) =&  \left( 1 - \frac{e^{-\tau}}{\Lambda} \right) e^{-L}L +   \frac{e^{-\tau}}{\Lambda}e^{-L}
   -\left(  1 - \frac{1}{\Lambda}  \right)e^{-L} L\\
=& \;  e^{-L} \left[  \frac{e^{-\tau}}{\Lambda}(1-L)  + \frac{L}{\Lambda} \right]  \\
=&\; e^{-\Lambda +\Lambda e^{-\tau}} \left[ \frac{e^{-\tau}}{\Lambda}(1-\Lambda( 1 - e^{-\tau}) ) + ( 1 - e^{-\tau})  \right]  \\
= & \;e^{-\Lambda +\Lambda e^{-\tau}} \left[   e^{-2\tau}  +  e^{-\tau} \left( \frac{1}{\Lambda} -2 \right)  + 1  \right].
\end{split}
\end{equation}
One can check that for $\tau = 0$, the above quantity reduces to $1/\Lambda$, as it should according to the resetting prescription (the term of order $1$ in $y$ in Eq. \ref{resetBoundary} is the inverse of the density in active boxes).
 Moreover, it goes to $e^{-\Lambda}$ when $\tau$ goes to infinity, which reflects the relaxation to a discrete Poisson distribution in the absence of resetting.\\

 Let us apply the adiabatic approximation to the system subject to resetting (at random times distributed exponentially in time $\tau$, with rate $r$).
 We can take the average of the variation rate of the density (Eq. \ref{densityDynA}) against  the
 time elapsed since the previous resetting time. The probability that this time equals $\tau$, up to $d\tau$, is $r e^{-r\tau}d\tau$ if the resetting rate $r$ is constant.
  The relaxation rate is therefore proportional to the resetting rate and to the Laplace transform of the density of active boxes at occupation 
 number $1$, taken at the resetting rate:
\begin{equation}\label{f1adiab}
\begin{split}
-\frac{d}{d\tau}\left( \frac{1}{\lambda(\tau)} \right)|_{\lambda = \Lambda} &= r \int_0^\infty e^{-r\tau} f_1(\tau) d\tau\\
& = r e^{-\Lambda}\int_0^\infty  d\tau e^{\Lambda e ^{-\tau}}\left[ e^{-(r+2)\tau} +  \left( \frac{1}{\Lambda} - 2  \right)e^{-(r+1)\tau}
   +  e^{-r\tau}  \right]\\
& = r \mathcal{L_\tau}\left[  f_1(\tau) \right] ( r ).
\end{split}
\end{equation}
When the resetting rate goes to zero, the value of the above product goes to the limit of the
 function $f_1$ when $\tau$ goes to infinity, which is $e^{-\Lambda}$. This is the prediction of the adiabatic approximation 
 in the ordinary case \cite{franz1995dynamical,godreche1995entropy}.

\begin{figure}
\includegraphics[width=0.98\textwidth]{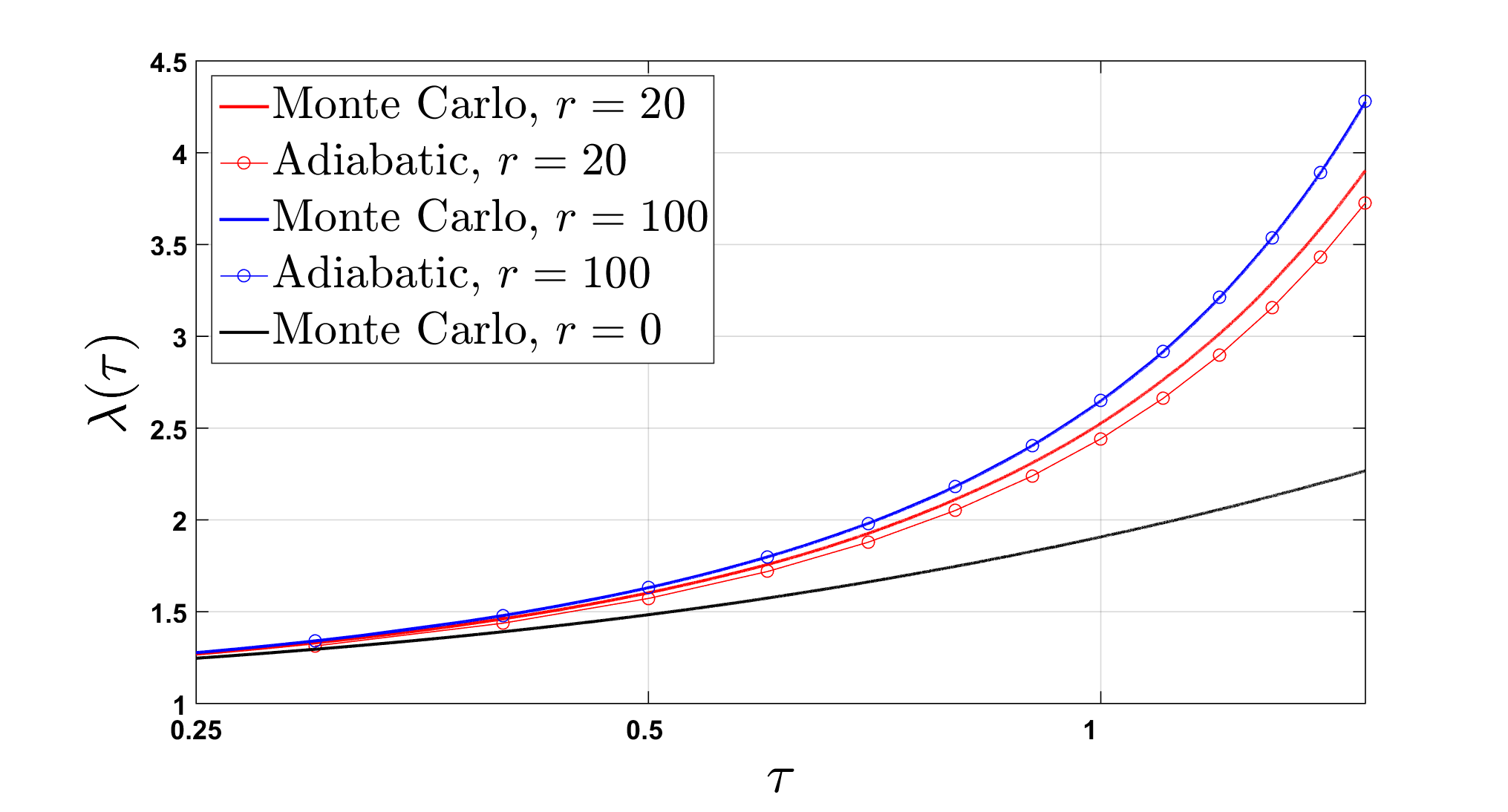}
 \caption{Density in active boxes {{(as a function of time $\tau$ plotted on a logarithmic scale)}}. 
 Number of boxes and particles  $M=N=10,000$. {{For $r=100$  (corresponding to $1\%$ probability of
 resetting at each Monte Carlo time step)}}
   the predictions obtained from the adiabatic approximation
 are all within $0.3\%$ of the Monte Carlo result.}
\label{figureModelA}
\end{figure}

\subsection{Numerical simulations}

 The dynamics of the model can be simulated directly by {{applying  the}}  microscopic rules
 to Monte Carlo samples: at each time step a particle is drawn uniformly from the system
 and put into a box drawn uniformly from the set of active boxes. The system is reset with 
 a constant resetting rate in time $\tau$, to a configuration in which  all active boxes except one contain just one
 particle. 
 At time $\tau = 0$, the density of active boxes with just one particle is $f_1(0) = 1$.
 During the first Monte Carlo step, the number of active boxes decreases by one unit.
 The variation rate of the density induced by Eq. \ref{densityDynA} during the first Monte Carlo step therefore reads:
\begin{equation}
 - \frac{1}{\Delta \tau} \Delta\left( \frac{1}{\lambda(0)}\right ) = - \frac{1}{\Delta \tau} \times\left( -\frac{1}{N}\right) = 1.
\end{equation}
 The Monte Carlo time step is therefore equal to the inverse of the size of the system:
 \begin{equation}
 \Delta \tau = \frac{1}{N}.
\end{equation}

On the other hand, {{the relaxation of the density can be predicted 
 by integrating  the adiabatic approximation numerically. Once the resetting rate $r$ has been chosen, the density is initialised at $\Lambda = 1$ (according to the initial configuration chosen in Eq. \ref{fromEq6}).
 The r.h.s.  of Eq. \ref{f1adiab}  is then evaluated (using the expression of the Laplace transform in terms of $\Lambda$ and $r$), and multiplied by the time step $\Delta \tau$.
 The resulting quantity is subtracted from the value of the inverse of the density. This induces a new value for the density,  $\Lambda$ is updated to this new value 
 and the process is iterated.
 The results of this numerical integration}} can be compared to the direct Monte Carlo simulation of the model (see Fig. \ref{figureModelA}
 for results at $N= M =10,000$). If the resetting is strong enough,  the validity of the 
 adiabatic approximation does not crucially depend on the density $\lambda$ being large enough (which is the case is the system without resetting,
 because the density of boxes containing just one particle becomes exponentially small when the density becomes large). At strong resetting,
 taking into account the short-time relaxation dynamics of the occupation-number probabilities, as in Eq. \ref{f1adiab}, is enough 
 to describe the relaxation of the system, as resetting probes short times, and there is only one particle moving at each Monte Carlo time step.


%
%
%
%

\section{Conclusion}
 In this paper we have considered the effect of resetting on two models with entropy 
 barriers. We used the known mapping of these models onto one-dimensional random walk problems with an absorbing trap at the origin,
  and the more recent results on diffusion with resetting. The proposed resetting prescriptions 
 allow for a prediction of the accelerated relaxation dynamics of the original  models under resetting. 
 At each resetting event, each non-empty box receives one particle, except one which receives the rest of the particles. 
  In the limit of a large system, this resetting configuration maps to a constant  resetting position (one) of the random walker.
  Resetting the system lowers the entropy barrier in the same way as resetting a random walker to its initial position 
 avoids wandering too far from the target. {{The entropy barrier corresponds to the fact that most of the states 
 in a given level set of energy are far from the boundary of the level set, when the energy is low enough. At later stages 
 of the relaxation, the system spends more and more time wandering in a fixed level set. The resetting process 
 lifts the entropy barrier (at constant energy) by putting the system on the boundary of the current level set, at stochastic times.}}\\

 In the case of the model with indistinguishable particles (termed model B in  \cite{godreche1995entropy}), 
 the corresponding random walk is symmetric, and upon a rescaling of time by the density in active boxes, 
 it becomes a simple diffusion with an absorbing trap at the origin. The exact results 
  of \cite{evans2011diffusion} imply that the model relaxes  in finite time, with a linear approach to the minimum energy.
 {{Resetting events accumulate towards the end of the process: in an experimental procedure, the rate of
 resetting in real time would have to be accelerated by continuously monitoring the energy of the system 
 (for developments on the experimental realisation of resetting prescriptions and their challenges, see \cite{experimentalResetting}).
 On a more general note, diffusion in a phase space consisting of many metastable states (corresponding 
 to configurations of a spin system) was shown to be logarithmic \cite{MezardPhaseSpace}. Time counted by number of moves in the system differs from real time:
   the typical lifetime of a state
 increases exponentially with the number of moves undergone by the system.
  For  recent developments on  the Glauber dynamics of spin systems in the Ising model under resetting, see \cite{IsingResetting}.}}\\

 In the case of the model with distinguishable particles (the zero-temperature version of the backgammon model, termed 
 model A in \cite{godreche1995entropy}), the random walk  is known to have position-dependent velocity and diffusion coefficient,
 which makes the explicit solution of the model much more difficult.  
    An adiabatic approximation can be applied, to estimate the relaxation of the density. This 
 approximation, introduced in \cite{ritort1995glassiness,godreche1995entropy}, has been adapted in this paper to the system with resetting 
 by working out the relaxation dynamics at short time scale. 
  On this time scale the occupation-number probability  is not in a quasi-stationary state, but the density can still 
 be treated as a constant.
 The time-evolution of the density has been expressed  in terms of the Laplace transform  (taken at the resetting rate),
 of the probability of occupation number one. This probability interpolates   between the inverse of the density (immediately
 after resetting)
 and the inverse exponential of the density (which is the ordinary case, reached
  after relaxation to a Poisson distribution of the occupation numbers).\\


\bibliography{bibRefsNew} 
\bibliographystyle{ieeetr}

\end{document}